# Magnetoresistance Relaxation in $(La_{0.5}Eu_{0.5})_{0.7}Pb_{0.3}MnO_3$ Single Crystals under the Action of a Pulse Magnetic Field

A.A. Bykov, S.I. Popkov, K.A. Shaykhutdinov, K.A. Sablina
Affilation: Kirensky Institute of Physics. Krasnoyarsk, Russia, 660036,
E-mail: alexey_bykov@iph.krasn.ru

## Abstract

Magnetoresistance of substituted lanthanum manganite $(La_{0.5}Eu_{0.5})_{0.7}Pb_{0.3}MnO_3$ in the pulse magnetic field H = 25 T was measured at different temperatures. Magnetoresistance relaxation with a characteristic time of $10^{-3}$ s was found. It has been established that the temperature dependence of the relaxation parameter τ(t) at different temperatures correlates with the temperature dependence of electrical resistance R(T). The proposed relaxation mechanism is related to relaxation of conducting and dielectric phases in the sample volume under phase stratification conditions. It is shown that relaxation parameter τ reflects the number of boundaries in the volume and not the ratio between phase fractions.

## Introduction

Substituted lanthanum manganites evoke great fundamental and practical interest due to the effect of colossal magnetoresistance (CMR) related to nanophase stratification in these materials [1]. One of the intriguing problem of the physics of manganites is relaxation of magnetoresistance and magnetization [2], which can elucidate the nature of physical mechanisms responsible for microscopic properties of a sample. In substituted lanthanum manganites, relaxation with two characterisitc times of about $10^{3}$ [3] и $10^{-3}$ s [2,4,5] was observed. The two main mechanisms responsible for magnetoresistance and relaxation in granular manganites work in different field ranges. The first is spin-dependent tunneling of carriers via dielectric spacers of the grains that, in their turn, can be magnetically ordered [6]. This mechanism usually determines magnetoresistance in weak (up to 1-10 kOe) fields when magnetization of ferromagnetic grains is not saturated. If dielectric boundaries are antiferromagnetically ordered, magnetoresistance, as well as relaxation and hysteresis, are observed in fields up to $10^5$ Oe. The second mechanism typical of substituted lanthanum manganites is classical negative magnetoresistance caused by nanophase stratification. Obviously, this magnetoresistance mechanism in its pure form can be experimentally observed in high-quality single-crystal films and bulk single crystals. In this study, we investigate the features of magnetoresistance and relaxation in strong pulse magnetic fields in the single-crystal substituted lanthanum manganite $(La_{0.5}Eu_{0.5})_{0.7}Pb_{0.3}MnO_3$.

## Experimental

The $(La_{0.5}Eu_{0.5})_{0.7}Pb_{0.3}MnO_3$ single-crystal sample was grown by spontaneous crystallization. PbO and $PbF_2$ were used as solvents. The sample was characterized in studies [7, 8]. Measurements in strong magnetic fields were performed on a pulse facility at the Laboratory of Strong Magnetic Fields, Kirensky Institute of Physics. The sample with attached contacts was placed on a detachable tip of the internal cryostat rod made of nonmagnetic ebonite with a built-in heat sensor. Transport characteristics (magnetoresistance and R(T)) were studied by the standard four-probe method. The signal from the sample was separated from the whole signal using digital lock-in amplifier. The absence of a backward half-wave was controlled by magnetic sensors. The measurements were performed on the parallelepiped sample $1\times0.5\times0.5$ $mm^3$ in size. Low-resistance contacts of conducting wires and the sample were made using an Epo-Tek H20-E contact adhesive.

## Results and discussion

Figure 1 shows temperature dependences of resistance R for the single-crystal $(La_{0.5}Eu_{0.5})_{0.7}Pb_{0.3}MnO_3$ sample measured in external dc magnetic fields of 0 – 90 kOe and in pulse magnetic fields up to 250 kOe (at this field value, the magnetoresistive effect completely saturates). The insert in the figure demonstrates the values of the magnetoresistive effect MR = ΔR(H)/R(0). For this sample, the temperature of the metal-dielectric transition at H = 0 Oe is $T_{max} \approx 185$ K. The magnetoresistive effect is observed over the entire temperature range 2 - 300 K. Figure 2 presents time dependences of a magnetic field and resistance of the sample. The moment of switching off the field can be clearly seen by the narrow peak at 15th millisecond of the pulse: this jump corresponds to closing the thyristor in the solenoid-capacitors oscillating circuit of the pulse facility. It can be seen that after switching off the field the resistance remains at a certain level and relaxes with time.

The authors who observed earlier the magnetoresistance relaxation with millisecond times in strong magnetic fields on polycrystalline films suggested [9,10] that the relaxation is related to the spin-dependent transport during carrier tunneling via antiferromagnetic boundaries of ferromagnetic grains of a manganite. However, the relaxation observed by us on the single-crystal sample cannot develop in this way, since there are no grain boundaries. Similar phenomena possibly occur at the twinning boundaries and defects in the crystal, but one can suggest that the relaxation is caused by the change in the ratio between conducting and dielectric phases in the crystal volume under the action of a magnetic field.

Figure 3 shows time dependences of magnetoresistance at different temperatures. The starting point of the curve coincides with the moment of switching off the field H = 250 kOe at which the magnetoresistive effect is maximum and saturated (Fig. 1). The narrow peak in the beginning is not related to magnetoresistance of the sample and is caused by an electromagnetic pulse that accompanies opening of the solenoid-power capacitors oscillating circuit. The obtained dependence R(t) can be easily approximated by the function $R(t)=R_0 exp(-t/T)$, where $R_0$ is the initial resistance and τ is the attenuation parameter. Parameters of the approximation were chosen by minimization of the discrepancy between the experiment and the function obtained. Then, the temperature dependence of parameter τ presented in Fig. 3 was built.

It can be seen that, qualitatively, the curve τ(T) follows the temperature dependence of resistance of the sample. Such a behavior of the curve can be explained with regard to the energy of boundaries of ferromagnetic regions («drops»). Upon cooling the sample, according to the phase stratification model, ferromagnetic regions or «drops» arise in the sample. Saturation of magnetoresistance in fields of about 250 kOe indicates that the entire sample volume is occupied by the ferromagnetic phase and there is no phase stratification. After switching off the magnetic field, the inverse process of phase stratification starts and the number of conducting and dielectric regions in the crystal is determined by temperature. At the temperature of the metal-dielectric transition, an infinite conducting cluster starts forming in the sample and the number of conducting and dielectric regions and, consequently, grain boundaries, is maximum. At these temperatures, we observe the maximum magnetoresistance relaxation. The degree of phase stratification and, correspondingly, parameter τ decrease with increasing distance from the temperature of the metal-dielectric transition towards both higher and lower temperatures.

**Conclusions**

Thus, the magnetoresistance relaxation with a characteristic time of $10^{-3}$ s in a wide temperature range has been found. It has been established that the temperature dependence of the relaxation parameter τ(T) is qualitatively consistent with the dependence R(T). It has been shown that parameter τ reflects the number of boundaries in a volume or the degree of phase stratification.

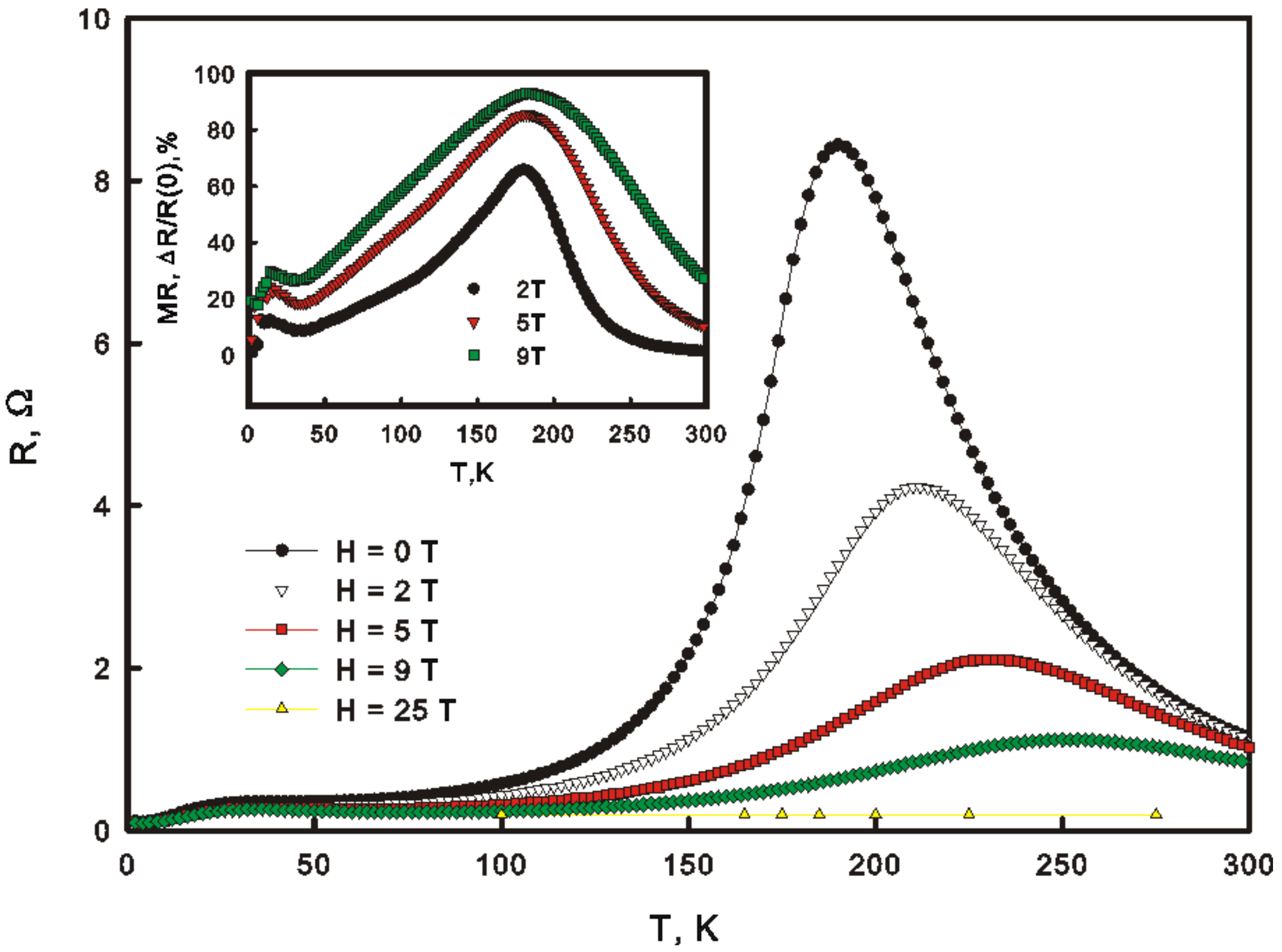

**Fig. 1. R(T) and ΔR/R0(T) dependences in different fields from study [8]. It can be seen that at H = 25 kOe magnetoresistance completely saturates.**

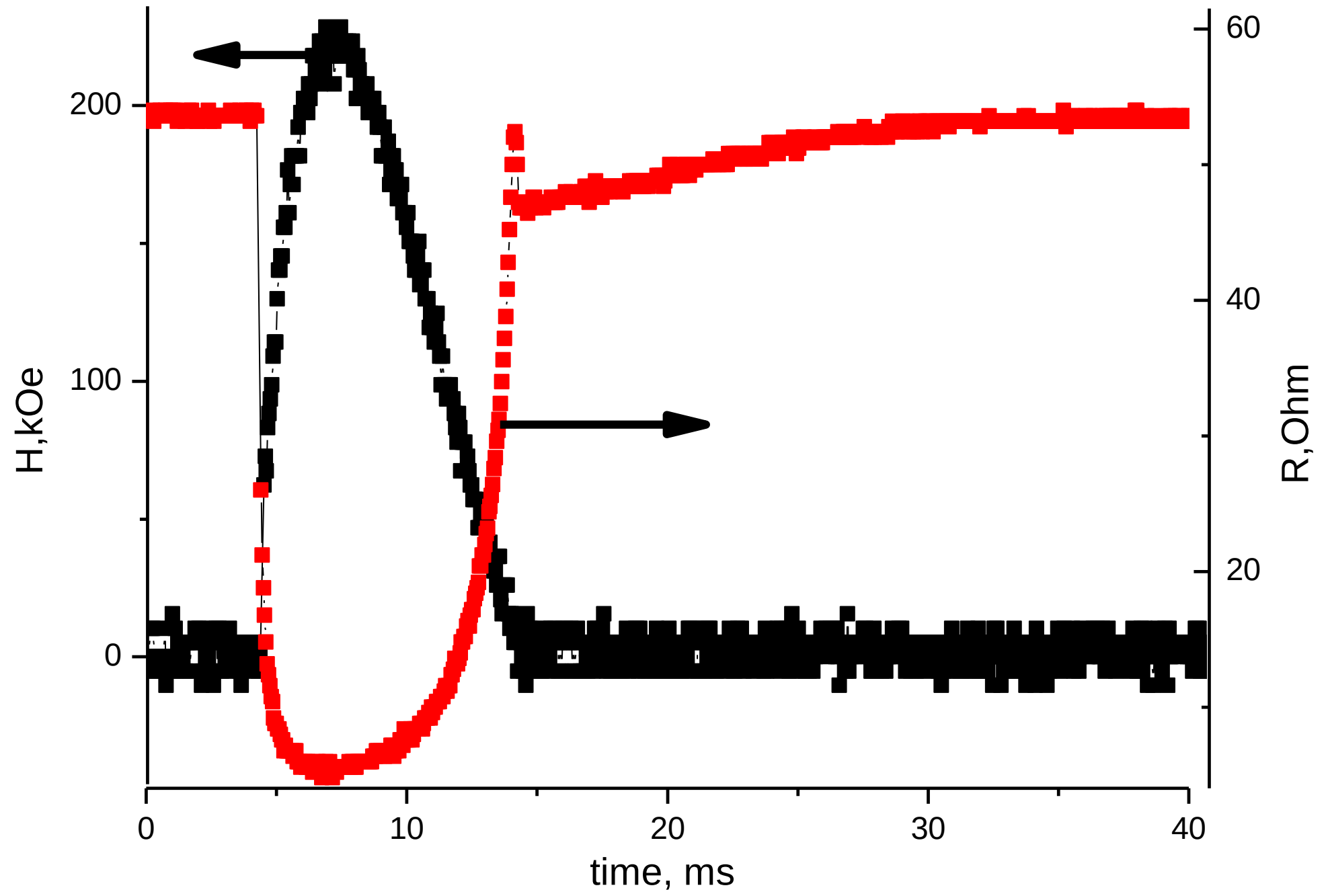

**Fig. 2. Time dependences of the applied field (black dots) and resistance (red dots). It can be seen that after switching off the field, the resistance remains at a certain level and relaxes in time.**

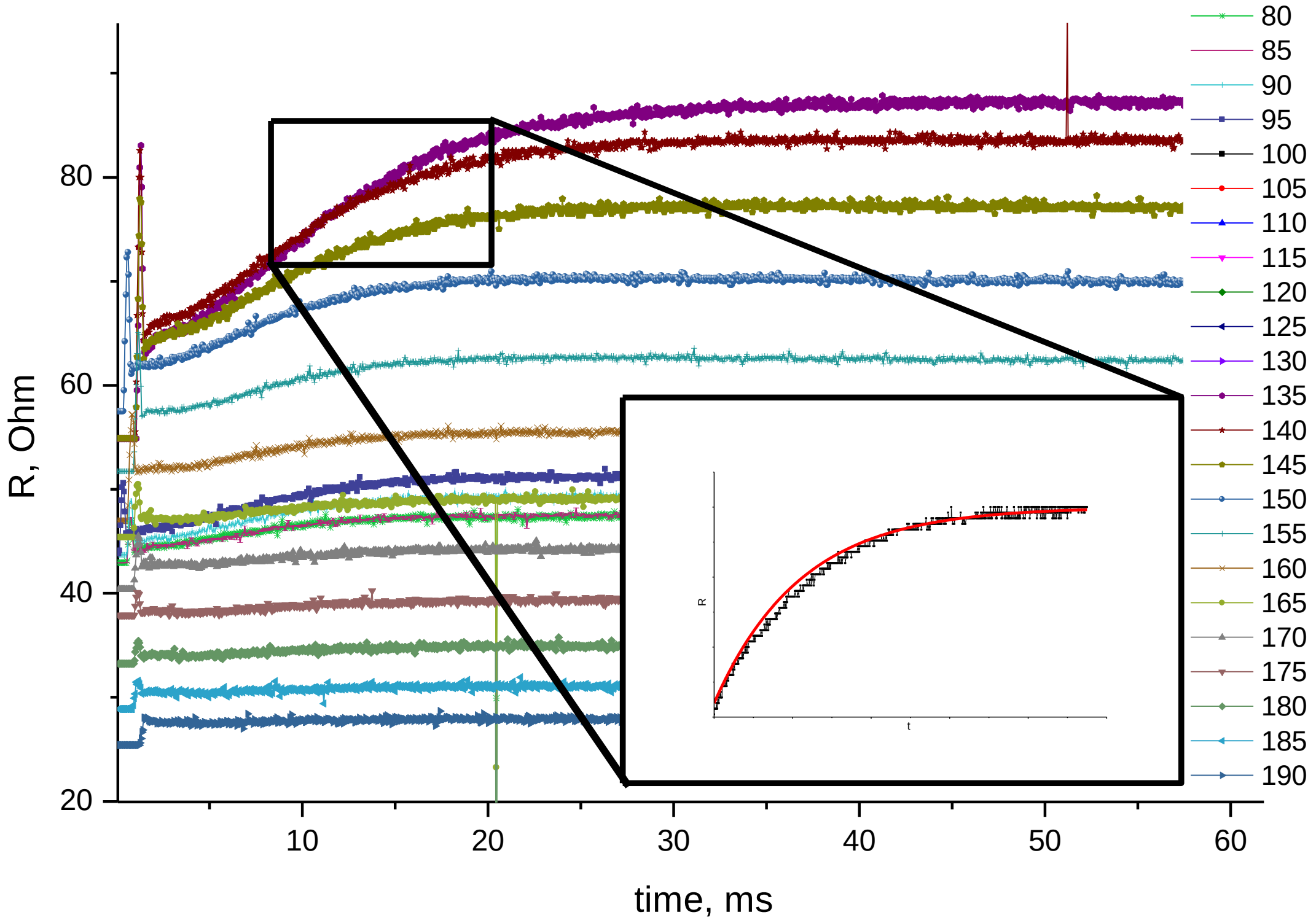

**Fig. 3. R(t) dependences at different temperatures and the field H = 250 kOe. Only the relaxation portion is presented: the curve part after switching off the magnetic field. Insert: R(T) dependence (black dots) at 175 K and approximation curve (red), with the parameter τ = 11 ms.**

.

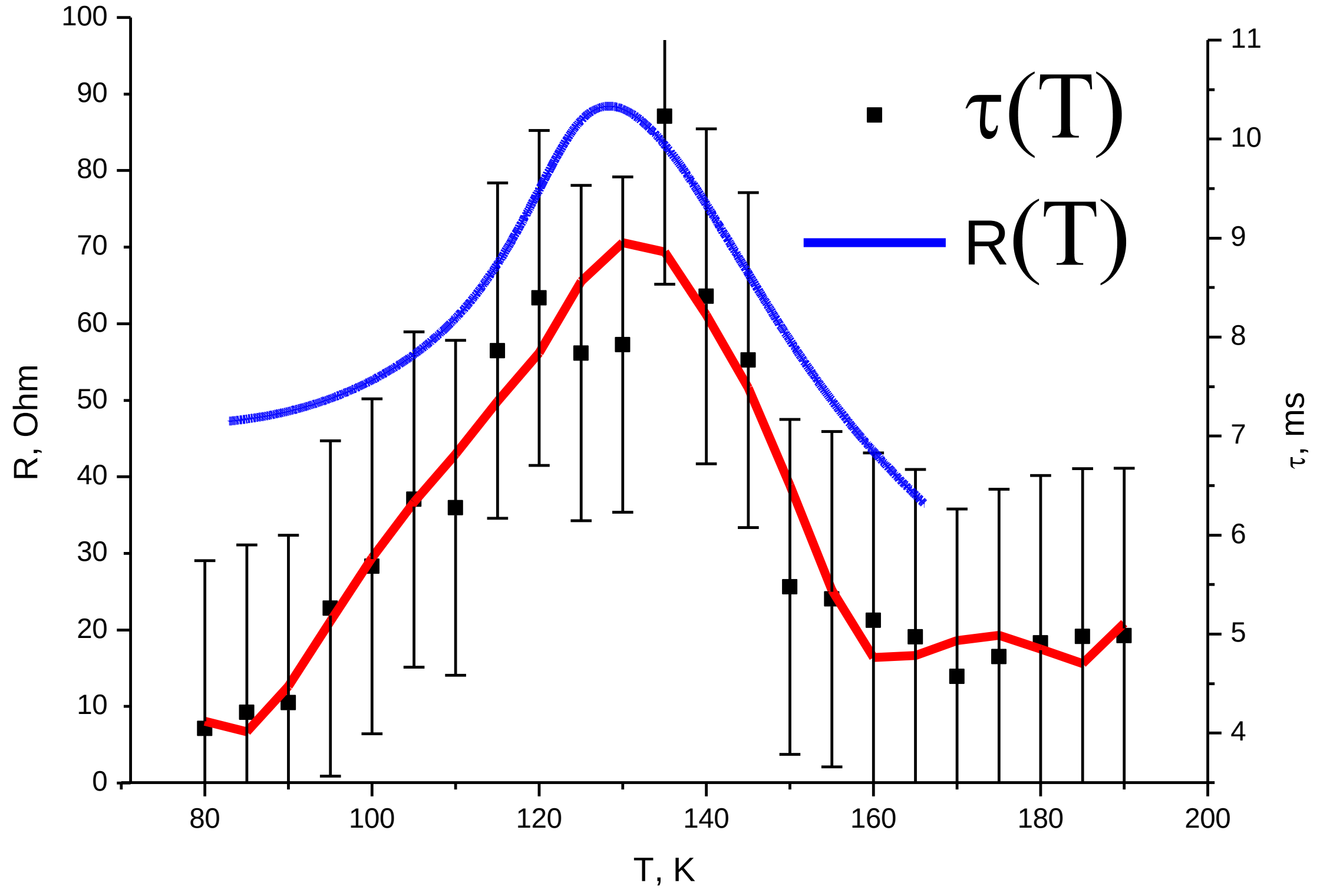

**Fig. 4. Temperature dependence of parameter τ (black dots), smoothed curve (red dots), and temperature dependence of resistance (blue line) for the same sample. It can be seen that the curve of parameter τ correlates with the resistance curve.**